\documentclass[l10pt,journal,compsoc]{IEEEtran}
\usepackage{amsmath,amsfonts}
\usepackage{algorithmic}
\usepackage{algorithm}
\usepackage{array}
\usepackage[caption=false,font=normalsize,labelfont=sf,textfont=sf]{subfig}
\usepackage{textcomp}
\usepackage{stfloats}
\usepackage{url}
\usepackage{verbatim}
\usepackage{graphicx}
\usepackage{cite}
\usepackage{xcolor}
\usepackage{multirow}
\usepackage{hyperref}
\usepackage{array} 
\begin{document}

\title{Automatic Configuration Tuning on Cloud Database: A Survey}





\author{Limeng Zhang and M. Ali Babar 
\IEEEcompsocitemizethanks{\IEEEcompsocthanksitem Limeng Zhang and M. Ali Babar are with the Centre for Research on Engineering Software Technologies (CREST), The University of Adelaide, Australia.\protect\\
E-mail: limeng.zhang@adelaide.edu.au \\
ali.babar@adelaide.edu.au 
}
\thanks{}}

\maketitle

\begin{abstract}
Faced with the challenges of big data, modern cloud database management systems are designed to efficiently store, organize, and retrieve data, supporting optimal performance, scalability, and reliability for complex data processing and analysis. However, achieving good performance in modern databases is non-trivial as they are notorious for having dozens of configurable knobs, such as hardware setup, software setup, database physical and logical design, etc., that control runtime behaviors and impact database performance. To find the optimal configuration for achieving optimal performance, extensive research has been conducted on automatic parameter tuning in DBMS. This paper provides a comprehensive survey of predominant configuration tuning techniques, including Bayesian optimization-based solutions, Neural network-based solutions, Reinforcement learning-based solutions, and Search-based solutions. Moreover, it investigates the fundamental aspects of parameter tuning pipeline, including tuning objective, workload characterization, feature pruning, knowledge from experience, configuration recommendation, and experimental settings. We highlight technique comparisons in each component, corresponding solutions, and introduce the experimental setting for performance evaluation. Finally, we conclude this paper and present future research opportunities. This paper aims to assist future researchers and practitioners in gaining a better understanding of automatic parameter tuning in cloud databases by providing state-of-the-art existing solutions, research directions, and evaluation benchmarks.

\end{abstract}


\section{Introduction}
In the increasingly digitized age, vast and diverse volumes of data are generated from various sources, including mobile devices, social media platforms, sensors, and more. Faced with this data explosion, cloud database management systems (DBMS) cloud database management systems (DBMS) for data storage, coupled with big data analytics frameworks (BDAF),  have emerged as powerful solutions to tackle the complexities of handling and processing massive and intricate data sets in a scalable and flexible manner. This makes them invaluable tools for organizations grappling with the challenges of big data and digital transformation~\cite{choi2014improving,sahatqija2018comparison}.

 However, achieving good performance in modern DBMSs is non-trivial. Modern DBMSs have hundreds of configurable knobs regarding hardware setup, software configuration, database physical and logical design, that affect their performance~\cite{zhang2022towards,yan2021workload,duan2009tuning,zhang2021restune}. Efficient parameter configurations can strike a balance between resource utilization, query responsiveness, and cost-effectiveness, while an inappropriate configuration can lead to significant performance degradation and inefficient usage of system resources~\cite{xin2022locat,van2017automatic,li2019qtune,zhu2017bestconfig,bao2018learning,zhang2019end,zhang2021restune,zhang2022towards,cai2022hunter}.

\subsection{Configuration Parameter Tuning Challenges}
In order to find an optimal configuration for a DBMS workload, either a database administrator or a tuning program has to address a number of challenges. First, \textit{the parameter space is vast and intricate}, with a large number of knobs existing in continuous space, and the number of DBMS knobs continues to grow with each new version and feature release. Staying abreast of these changes and comprehending their implications for system performance presents an ongoing challenge for database administrators~\cite{van2017automatic,zhang2022towards}. Second, \textit{the  interdependence of parameters within DBMSs complicates the optimization process}. Altering one knob may have repercussions on the effectiveness of another, introducing nonlinear correlations that pose difficulties for both manual and model-based configuration identification. Third, \textit{the diversity and heterogeneity of workloads, hardware, and other environmental factors in a DBMS further contribute to the complexity}. A DBMS may host numerous instances, each accommodating various types of workloads. Real-world application workloads are dynamic, with properties such as workload types, workload parallelization, workload records, and operations counts exhibiting variations over time~\cite{zhang2022towards, li2019qtune}. Furthermore, \textit{training samples are scarce} due to the time-intensive process of gathering historical observations and the complexity introduced by high-dimensional parameters. Addressing these challenges requires sophisticated approaches and tools to enhance the efficiency of DBMSs in diverse and dynamic operational environments.

\subsection{Contributions}
Through an investigation into the state-of-the-art automatic DBMS configuration tuning, we aim to provide researchers, developers, and practitioners with comprehensive insights and guidance in the field. In this study, our contributions are fourfold:
\begin{enumerate}
    \item We provide an overview of a wide spectrum of prominent automatic tuning methods for cloud databases, as discussed in various studies~\cite{zhu2017bestconfig,van2017automatic,van2021inquiry,bao2018learning,li2019qtune,zhang2019end,tan2019ibtune,zhang2021restune,zhang2022towards,cai2022hunter,kanellis2022llamatune,cereda2021cgptuner,trummer2022db}. These methods include Bayesian optimization-based methods, Neural network-based methods, Reinforcement learning-based methods, and Search-based methods. Meanwhile, we also include several relevant automatic tuning methods~\cite{kunjir2020black,song2021spark,xin2022locat,lin2022adaptive}, which focus on big data analytics framework but encounter similar challenges in identifying the optimal configuration within the complex and dependent configuration space.
    \item We outline the primary tuning objectives and summarize three main constraints or factors frequently discussed in automatic configuration tuning in DBMS (Section~\ref{tuningObjective}): overhead, adaptivity, and safety.
    \item We provide a comprehensive overview of the automatic tuning pipeline, encompassing Workload Characterization (Section~\ref{secEmbedding}), Feature Pruning (Section~\ref{SecFeatureSelection}), Knowledge from Experience (Section~\ref{SecKnowledgeExtraction}), Configuration Recommendation (Section~\ref{SecRecomendation}). We describe in detail the key features in each part.
    \item We providing a comprehensive summary of popular open-source benchmarks utilized for evaluating database performance optimization.
\end{enumerate}

\begin{table*}[htbp]
  \centering
   \caption{\label{TuningObjective}Summary of tuning objective} 
  \resizebox{\linewidth}{!}{
    \begin{tabular}{|c|p{2cm}|c|c|p{3cm}|p{1.3cm}|p{2cm}|p{2.5cm}|p{2cm}|}
    \hline
    \multicolumn{1}{|c}{\multirow{2}[0]{*}{\textbf{Category}}} & \multicolumn{1}{|c}{\multirow{2}[0]{*}{\textbf{Work}}} & \multicolumn{2}{|c}{\textbf{Target}} & \multicolumn{2}{|c}{\textbf{Objective}} & \multicolumn{3}{|c|}{\textbf{Constraints/Factors}}   \\ \cline{3-9}
          &       & \multicolumn{1}{c|}{\textbf{DBMS}} & \multicolumn{1}{c|}{\textbf{BDAF}} & \multicolumn{1}{c|}{\textbf{Performance}} & \multicolumn{1}{c|}{\textbf{Others}} & \multicolumn{1}{c|}{\textbf{Overhead}} & \multicolumn{1}{c|}{\textbf{Adaptivity}} & \multicolumn{1}{c|}{\textbf{Safety}} \\ \hline
    \multirow{7}[0]{*}{BO} & Ottertune~\cite{van2017automatic} & \checkmark & -     & latency & -     & -     & -     & - \\ \cline{2-9}
          & RelM~\cite{kunjir2020black}&       & \checkmark & execution time & memory efficiency & memory allocation efficiency & -     & out of memory \\ \cline{2-9}
          & RestTune~\cite{zhang2021restune} & \checkmark & -     &       & resource utilization & CPU, I/O and memory & -     & - \\ \cline{2-9}
          & CGPTuner~\cite{cereda2021cgptuner} & \checkmark & -     & throughput, response time, etc. & -     & -     &  JVM and OS & - \\ \cline{2-9}
          & LlamaTune~\cite{kanellis2022llamatune} & \checkmark & -     & throughput,  latency & -     & -     & -     & - \\ \cline{2-9}
          & ONLINETUNE~\cite{zhang2022towards} & \checkmark & -     & throughput, latency, etc. & -     & -     & underlying data & performance degradation \\ \cline{2-9}
          & LOCAT~\cite{xin2022locat} & -     & \checkmark & execution time & -     & -     & data size & - \\ \hline
    \multirow{4}[0]{*}{RF} & CDBTune~\cite{zhang2019end} & \checkmark & -     & throughput and  latency & -     & -     & -     & - \\ \cline{2-9}
          & Qtune~\cite{li2019qtune} & \checkmark & -     & throughput and latency & -     & -     & -     & - \\ \cline{2-9}
          &MMOT~\cite{gur2021adaptive}& \checkmark & -     &throughput and latency & -     &   -  & -  & - \\ \cline{2-9}
          &WATuning~\cite{ge2021watuning}& \checkmark & -     &throughput and latency & -     &   -  & read/write ratio & - \\ \cline{2-9}

          & HUNTER~\cite{cai2022hunter} & \checkmark & -     & throughput and  latency & -     & -     & -     & - \\ \cline{2-9}
          &DB-BERT~\cite{trummer2022db} & \checkmark & -     & throughput or execution time & -    & -     & -     & - \\ \hline     
     \multirow{1}[0]{*}{NN}  & LITE~\cite{lin2022adaptive} & -     & \checkmark  & latency & -     & -     & application diversity and cluster environment & - \\ \cline{2-9}

    & iBTune~\cite{tan2019ibtune} & \checkmark     & -  & - & buffer size    & -     & - & - \\ \hline
    
    \multirow{3}[0]{*}{Search} & BestConfig~\cite{zhu2017bestconfig} & \checkmark & \checkmark & user-defined performance & -     & -     & -     & - \\ \cline{2-9}
          & AutoTune~\cite{bao2018learning} & -     & \checkmark & execution time & -     & CPU, memory, disk and network & -     & - \\ \cline{2-9}
          & UDAO~\cite{song2021spark} & -     & \checkmark & execution time & -     & user-concerned aspects & -     & - \\ \hline
    \end{tabular}%
    }
  \label{tab:addlabel}%
\end{table*}%

\section{Tuning Objectives}\label{tuningObjective} 

In the context of DBMS configration uning, the primary objective is to maximize the \textit{Performance}, such as throughput and latency (e.g., 95th percentile
latency). This aims to gauge the DBMS's ability to efficiently handle a higher volume of workloads or queries.

Additionally, several other aspects/objectives can also be considered during the tuning process as summarised in Table~\ref{TuningObjective}, including:

\begin{itemize}

    \item \textit{Overhead}: Focuses on the amount of time or system resources the method requires to recommend knob settings, including tuning time, Total Cost of Ownership (TCO), and resource utilization (such as CPU utilization), etc. For example, RelM~\cite{kunjir2020black} concentrates on optimizing memory allocation for such applications. RestTune~\cite{zhang2021restune} aims to reduce resource utilization while still guaranteeing the Service Level Agreement (SLA), e.g., without violating throughput and latency requirements.

    \item \textit{Adaptivity}: Evaluates how well the method performs in new and varying scenarios, such as hardware adaptation for changes in I/O, memory, workload adaptation for varying request rates and data sizes, etc.

    \item \textit{Safety}: One crucial facet of safety lies in the ability of a tuning method to avoid recommending parameter configurations that could potentially degrade system performance. ONLINETUNE~\cite{zhang2022towards}, which focuses on tuning online databases safely in changing cloud environments, addresses this concern. The safety threshold, as defined in ONLINETUNE, is anchored to the baseline performance exhibited under the default configuration, denoted as default performance. Beyond performance-centric safety assessments, resource allocation represents another pivotal dimension. Within this context, Kunjir et al.~\cite{kunjir2020black} present RelM, wherein safety concerns are framed in terms of resource utilization aligning with allocated thresholds

\end{itemize}

\section{Overview of Tuning Framework}\label{secOverview} 

\begin{figure*}[htbp]
    \centering
    \includegraphics[width=\textwidth]{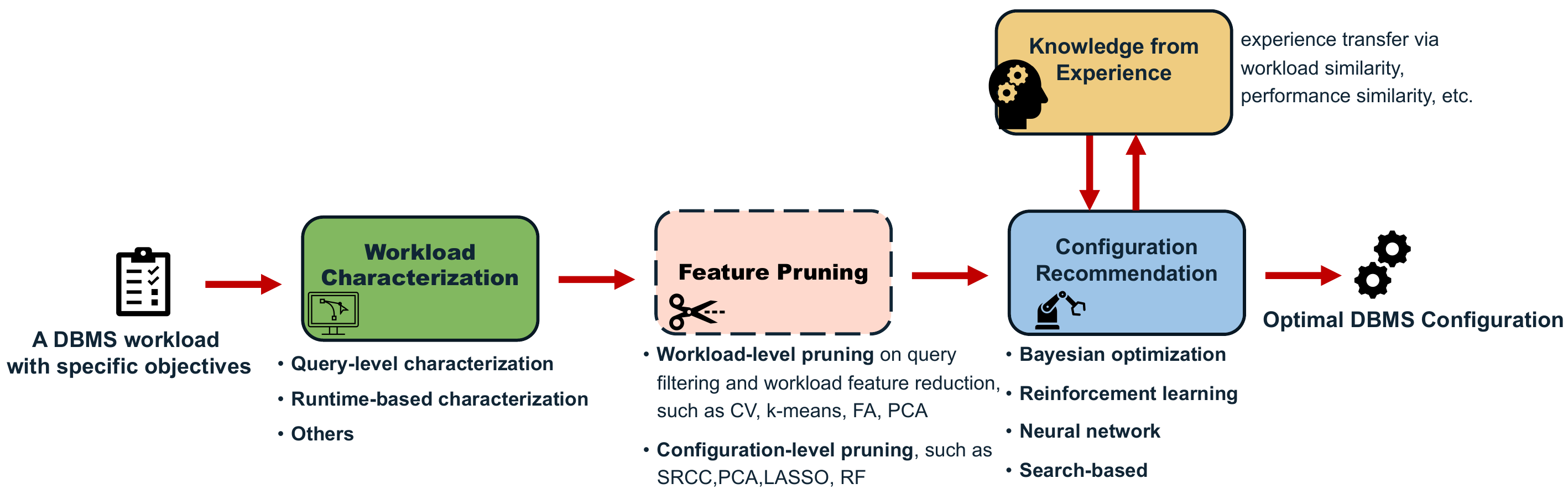}
    \caption{The general framework of automatic configuration tuning on DBMS. }
    \label{fig:overview}
\end{figure*}

In this study, we break down the entire knob tuning pipeline into five essential components: \textit{Workload Characterization, Feature Pruning, Knowledge from Experience, Configuration Recommendation} as depicted in Fig.~\ref{fig:overview}. The initial phase, \textit{Workload Characterization}, is crafted to comprehensively model a workload, utilizing either logical-level queries or embedding runtime metrics for a dynamic understanding of performance indicators \textit{Feature Pruning}, applied both at the workload-level to minimize execution time and at the configuration-level to streamline the search space, is strategically employed to enhance efficiency through the application of pruning techniques.  \textit{Knowledge from Experience} draws upon historical insights, enabling the tuning algorithm to efficiently converge towards optimal configurations.  Finally, \textit{Configuration Recommendation} involves utilizing diverse techniques, including Bayesian Optimization, Neural networks, Reinforcement learning, and Search-based methods, to generate configurations finely tailored to the specific characteristics of the workload. 

\section{Workload Characterization}\label{secEmbedding}
The initial step in the tuning system involves modeling the  characteristics of the target workload to facilitate the tuning algorithm in acquiring knowledge from historical data, thereby accelerating the tuning process or improving coverage to achieve optimal configurations. In this section, we present two main aspects for modeling DBMS workloads: a query-level characterization based on the queries composed in a workload and a runtime-based characterization based on DBMS runtime statistics obtained during the execution of the instance.

\subsection{Query-level Characterization}
Query-level characterization aims at modeling a workload from its composed queries~\cite{zhang2022towards}. A DBMS application allows users and other applications to engage with the database by formulating and executing queries, such as SELECT, INSERT, UPDATE, DELETE—tailored to manipulate and retrieve information and accomplish various tasks. Hence, researchers have endeavored to model a workload by analyzing the composition of queries and utilizing query characteristics to represent the workload~\cite{zhang2022towards}. Existing work on query-level characterization can be summarized as follows:

\begin{itemize}  
   \item \textit{Features from the query text:} In a DBMS workload, each query will perform a certain type of operation within the DBMS. Hence, some researchers try to characterize a workload by analyzing its static query text and then merging each query vector to ultimately represent the workload. RestTune~\cite{zhang2021restune} analyzes the cost of a MySQL workload by examining the distribution of queries within each workload based on their reserved SQL keywords (e.g., SELECT, UPDATE, DISTINCT). Initially, it identifies the reserved operation keywords for each query and subsequently computes their TF-IDF (term frequency-inverse document frequency) values. These TF-IDF feature vectors serve as input for a random forest model utilized for classification. This model provides a predicted probability distribution regarding predefined resource cost levels for each query. Ultimately, the average of the probability distributions for all queries across the entire workload is calculated to represent the meta-feature cost for the input workload. LOCAT~\cite{xin2022locat} features a Spark SQL workload directly with its query composition. It then filters out some unimportant queries and uses those important queries to represent a workload when evaluating the performance of recommended configuration. ONLINETUNE~\cite{zhang2022towards} features a workload in terms of the query arrival rate within it and its query composition. It leverages a standard LSTM encoder-decoder network to encode each query in the workload. Subsequently, it averages the query encodings to obtain the composition feature of the workload.
   
\item \textit{Features from the query plan:} In addition to the static query text, some researchers model a query through looking into its execution process.  When a query is submitted to the DBMS, the system first parses the query to check its syntax and semantics. After parsing, the query is compiled into an execution plan, which outlines the steps needed to execute the query efficiently. Some researchers represent a query  through parsing its query plan to aggregate key features, including the cost or categories of operators, scanned tables, and predicates, and then featurizing the workload with the query vectors. In Qtune~\cite{li2019qtune}, each SQL query is represented with features including query type (e.g., insert, delete, select, update), involved tables, and the estimated processing cost derived from inherent operations (e.g., scan, hash join, aggregate). Then, it merges all the query vectors into a unified workload vector according to its predefined merging strategies for each part.
  
  \item \textit{Features from other query-related factors.} In addition to the aforementioned query features, certain studies consider the underlying data distribution in the database as an informative factor, based on the observation that only data changes affecting the workload queries are relevant to the tuning policy. Building upon this premise, ONLINETUNE~\cite{zhang2022towards} extracts the database data distribution by leveraging insights from the DBMS optimizer, which encompass estimates of rows scrutinized by queries, the proportion of rows filtered by table conditions in queries, and the utilization of indexes. Subsequently, it computes the average of these three query features to derive the underlying data distribution associated with a workload. Finally, combined with the features extracted from query text, it concatenates the workload feature and underlying data feature to obtain the final contextual features.

\end{itemize}

\subsection{Runtime-based Characterization} 
In addition to characterizing a workload in terms of queries, a workload can also be characterized by its runtime characteristics. Modern DBMSs provide extensive information about workload running behavior~\cite{van2017automatic}. For example, MySQL’s InnoDB engine~\cite{InnoBDMetrics} provides statistics on the number of pages read/written, query cache utilization, and locking overhead. OtterTune~\cite{van2017automatic} characterizes a workload using such numeric metrics, such as the number of pages read/written, query cache utilization, and locking overhead, etc, to reflect various aspects of its runtime behavior. Additionally, researchers can also define new performance indicators tailored to unique requirements. For instance, RelM~\cite{kunjir2020black} profiles the memory allocation of a BDAF application with different configuration parameters, along with customized performance metrics regarding memory management decisions at multiple levels, including the resource-management level, container level, application level, and inside the Java Virtual Machine.

Apart from workload-related and runtime-related features, other factors associated with workload execution can also be considered, such as features related to running experiments and data. LITE~\cite{lin2022adaptive} provides insights on incorporating code semantics features and scheduler features alongside model input data features (such as column number, rows, iteration number, and partition number), as well as cluster environment factors related to nodes, memory, CPU frequency, and bandwidth when generating configurations for applications in Spark.

\section{Feature Pruning}\label{SecFeatureSelection}

Given the diversity of workloads and the high-dimensional configurable parameters of DBMSs, both data collection and configuration space search can be time-intensive. Hence, employing strategies to mitigate workload execution time and streamline the search space becomes imperative. In this section, we present pruning strategies at two levels: workload-level pruning to optimize workload execution time and configuration-level pruning for enhancing search space efficiency.

Table~\ref{FeaturePruningAndKnowledge} provides a technical comparison of strategies on feature pruning (in Section~\ref{SecFeatureSelection}) and knowledge extraction (in Section~\ref{SecKnowledgeExtraction}) from existing tuning methods.

\begin{table*}[htbp]
  \centering
  \caption{\label{FeaturePruningAndKnowledge} Strategies of Feature Pruning and Knowledge From Experience}
    \begin{tabular}{|l|l|l|p{4cm}|p{4cm}|}
    \hline
    \multicolumn{1}{|c}{\multirow{2}[0]{*}{\textbf{Work}}} & \multicolumn{2}{|c}{\textbf{Feature Pruning}} & \multicolumn{2}{|c|}{\textbf{Knowledge From Experience}} \\  \cline{2-5}
          & \multicolumn{1}{l|}{\textbf{Workload-level}} & \multicolumn{1}{l|}{\textbf{Configuration-level}} & \multicolumn{1}{c|}{\textbf{Strategy}} & \multicolumn{1}{c|}{\textbf{Aims}} \\ \hline
          LOCAT~\cite{xin2022locat}& CV&SRCC, KPCA       & - &\\ \hline
          Ottertune~\cite{van2017automatic}&  FA, k-means& LASSO      & Euclidean distance& Training data construction\\\hline
          ResTune~\cite{zhang2021restune}& - &- &Epanechnikov quadratic kernel (static), Performance surface (dynamic)& Model ensembling weight\\\hline
          LlamaTune~\cite{kanellis2022llamatune}& - & HeSBO&-&-\\\hline
          HUNTER~\cite{cai2022hunter} &PCA&RF&-&-\\\hline
          
          CGPTuner~\cite{cereda2021cgptuner}& - &- & -& Data normalization, Model training\\ \hline
           MMOT~\cite{gur2021adaptive}& 3-layer autoencoder&-  & Cosine similarity &Model selection\\

           \hline
    \end{tabular}%
  \label{tab:addlabel}%
\end{table*}%

\subsection{Workload-level Pruning}

Workload-level pruning strategies primarily contribute to two aspects: data collection and model training. \textit{Query-level filtering} aims to remove unimportant queries from a workload, thereby reducing the execution time of each workload and accelerating both training data collection and model validation. \textit{Workload Feature Selection} aims to reduce the dimensionality of each workload feature, thereby decreasing the training time for autotuning models.

\subsubsection{Query-level Filtering}

Considering that each application  consists of a number of queries, the execution time of a workload can be decreased through removing some unimportant queries. Query filtering aims to removing the unimportant queries which are not sensitive to the configuration parameters.
\begin{itemize}
    \item Coefficient of Variation (CV): is a statistical measure that provides insights into the relative variability of a set of values. It is calculated as the ratio of the standard deviation to the mean, expressed as a percentage. LOCAT~\cite{xin2022locat} utilizes CV to identify and eliminate configuration-insensitive queries whose performance (query execution time) is not affected by configuration changes within an application. It first constructs a matrix of the execution time for each query under different Spark SQL configurations (30 configuration instances) and then calculates the CV to filter queries with smaller CV values, categorizing them as 'configuration-insensitive'.
    
\end{itemize}

\subsubsection{Workload Feature Selection}
Workload feature selection aims to reduce the dimensionality of workload features. Existing strategies applied in the automatic parameter tuning on DBMSs include:
\begin{itemize}
    \item k-means: is a popular unsupervised machine learning algorithm used for partitioning a dataset into clusters based on similarity. The primary goal of k-means is to group data points in such a way that points within the same cluster are more similar to each other than to points in other clusters. In Ottertune \cite{van2017automatic}, each workload is represented with runtime metrics. When conducting workload feature selection, \textit{factor analysis (FA)} is first applied to transform the (potentially) high-dimensional DBMS metric data into lower-dimensional features. Then \textit{k-means}, to cluster this lower dimensional data into meaningful clusters (for reducing noise data)  and select  a single metric for each cluster, namely, the one closest to the cluster center. During the process,  number of clusters is determined through a heuristic model proposed in the work \cite{hastie2009elements}. 

    \item Factor analysis (FA):  is a statistical technique used for identifying underlying latent factors that explain the observed relationships among a set of variables. The key assumption in factor analysis is that observed variables are linear combinations of unobservable latent factors and random error terms.  Ottertune \cite{van2017automatic} utilizes it to transform the (potentially) high-dimensional DBMS metric embedding vector into lower-dimensional features. 
    
    \item Principal Component Analysis (PCA):  is a popular linear  dimension reduction method that transforms correlated variables into uncorrelated ones, known as principal components. It does so by calculating the covariance matrix, performing eigendecomposition, and selecting a subset of principal components based on their variance. HUNTER~\cite{cai2022hunter} applies PCA to complete the  feature reduction on its collected workload runtime metrics for a workload, which including number of lush statements executed, number of threads activated, etc.

    \item Autoencoder: is a neural network architecture meticulously devised to achieve precise reconstruction of its input. Comprising two integral components, namely the encoder and the decoder, it operates by first directing the input through the encoder to produce a latent code. This latent code subsequently serves as the foundation for the decoder to generate the output. Thus, the latent code assumes a pivotal role in the identification and disentanglement of various sources of variation inherent within the input data. Through the reduction of the latent code's dimension, the autoencoder facilitates effective dimension reduction, thereby yielding a more condensed and meaningful representation of the input data. MMOT~\cite{gur2021adaptive} utilizes the autoencoder architecture to extract inherent patterns among workloads and then uses these patterns to identify the model from similar workloads to generate configurations.
    
\end{itemize}

\subsection{Configuration-level Pruning}
Another approach to mitigating the optimization time required by machine learning-based methodologies is to reduce  configuration space necessitating tuning. Configuration-level pruning endeavors to discern and prioritize the tuning of parameters that exert substantial influence on the DBMSs  based on historical observations.

\begin{itemize}
    \item Spearman Rank Correlation Coefficient (SRCC) ~\cite{zar2005spearman}:  is a nonparametric measure of rank correlation on the strength of a monotonic relationship between paired data. It is defined as the Pearson correlation coefficient between the rank variables. By leveraging its capacity to gauge monotonic relationships, SRCC facilitates the identification and removal of redundant parameters, contributing to a more efficient and simplified configuration. LOCAT~\cite{xin2022locat} calculates \textit{SRCC} between each configuration parameter and the workload execution time to filter out some redundant configuration parameters and narrow down the parameter search space."

    \item Principal Component Analysis (PCA):  is a dimensionality reduction method that transforms correlated variables into uncorrelated ones, known as principal components. It does so by calculating the covariance matrix, performing eigendecomposition, and selecting a subset of principal components based on their variance. After SRCC, LOCAT ~\cite{xin2022locat} applies \textit{Gaussian kernel Principal Component Analysis (KPCA)} to further extract and form new lower-dimensional parameters derived after removal of redundant  configuration parameters for subsequent optimal parameter searching. 

    \item Least Absolute Shrinkage and Selection Operator (LASSO) and Lasso Path Algorithm. In the context of linear regression, LASSO modifies this objective function by adding a $\mathit{L_{1}}$ penalty term equals to $\lambda$ times the sum of  absolute values of the regression coefficients, where $\lambda$  controls the strength of the penalty, and it helps to reduce the effect of irrelevant variables  by penalizing models with large weights. A higher $\lambda$ value results in more coefficients being pushed toward zero, thereby achieving interpretable, stable, and computationally efficient feature selection. OtterTune~\cite{van2017automatic} uses the \textit{LASSO path algorithm} to assess the importance of configuration parameters by gradually reducing the penalty term. The order in which parameters are included during this process is then considered as the importance order, and OtterTune uses this information to prune less important parameters.
    
    \item Random Forest (RF): is a supervised machine learning method to deal with classification and regression problems. It can be deemed a voting algorithm consisting of multiple decision trees, used to calculate the importance of features. In the study~\cite{kanellis2020too}, the authors  utilize Latin Hypercube Sampling (LHS) to select a subset of configurations. Subsequently, they employ Random Forests with 300 Classification and Regression Trees (CART) to ascertain the importance of each parameter. Their findings reveal that when employing the YCSB workload-A on Cassandra, optimizing just five parameters can yield performance levels comparable to 99\% of the best configurations attained through extensive parameter tuning. HUNTER~\cite{cai2022hunter} adopt the RF classifier with CARTs and adopts the majority voting principle to  determine the importance of the knobs. It then selects the top-20 knobs to reduce the configuration search space in its tuning framework.
    
    \item    Hashing-enhanced Subspace BO (HeSBO)~\cite{nayebi2019framework}: is a randomized low-dimensional projection approach that operates by deducing the search space for Bayesian Optimization-based solutions (discussed in Section~\ref{SecBO}). By utilizing two uniform hash functions to construct the rows and signs of the random projection matrix, it performs a count-sketch projection to transform the original high-dimensional configuration space into lower-dimensional subspace. This process is intuitively designed to effectively preserve the characteristics, such as pairwise point distances, of the original high-dimensional space for the reduced (important) dimensions. LlamaTune~\cite{kanellis2022llamatune} adopts it to perform a projection of the configuration space from all documentations to a lower-dimensional subspace  and then use the Bayesian Optimization-based optimizer to tune this smaller subspace.

    \item Alternative approaches: Due to the challenges of tuning in high-dimensional configuration spaces, the identification of parameters that significantly affect DBMS performance has gained considerable attention. Among them,  Researchers in~\cite{cao2020carver} propose Caver to  select a subset of storage parameters. Given a set of initial parameters and configurations, Carver first employs LHS to select a small number of configurations for evaluation. Subsequently, it greedily selects the most important parameters based on their parameter importance values. The first parameter is determined by the highest importance value, after which Carver fixes its value and computes the parameter importance  values for the remaining parameters. During this process, importance is calculated as the difference between the variance of the original set of configurations and the sum of variances within the subset when parameters take different values. The algorithm selects parameters with the maximum difference.  
\end{itemize}

Recently, OpAdviser~\cite{zhang2023efficient} proposed constructing an effective configuration region that contains promising configurations. It leverages performance similarity among historical workloads to guide the construction process. Specifically, it extracts a relatively compact region based on similar workloads and a larger region based on less similar ones. Furthermore, OpAdviser constructs the target search space by employing a majority weighted voting strategy to aggregate suggestions from candidate tasks.

\subsection{Summary}
Given the complex configuration space and the diversity of workloads, employing pruning techniques to reduce workload running time and configuration search space emerges as a natural approach in addressing these complexities. In this section, we aim to provide direction for future practitioners and researchers on improving data collection efficiency and training efficiency through various pruning strategies. Specifically, we classify the pruning techniques into two levels: workload-level and configuration space. Regarding the workload level, we categorize it into two directions: eliminating redundant queries and workload feature reduction techniques. For the configuration level, we present the existing feature reduction methods applied in the state-of-the-art tuning methods, mainly focusing on feature projection, importance ranking, or feature clustering. 

Moreover, future researchers and practitioners can also try to explore different dimensionality reduction techniques tailored to specific data characteristics, as outlined in the survey by Hou et al.~\cite{hou2022dimensionality}. Furthermore, advancements in high-dimensional data technology offer opportunities to enhance the performance of feature pruning methods. For instance, Yang et al.~\cite{yang2019lasso} proposed an innovative variant of LASSO, named Efficient Tuning of Lasso (ET-Lasso), which focuses on ensuring feature selection consistency. This method has demonstrated effectiveness in efficiently selecting active features contributing to the response, achieved by integrating permuted features as pseudo-features within linear models.

\section{Knowledge from Experience}\label{SecKnowledgeExtraction}

As the process of finding the optimal configuration can be complex and time-consuming, a natural and common approach is to draw insights from previous workloads with similar characteristics and leverage that experience to guide the tuning algorithm in configuring the target workload. Existing aspects can be considered for learning from historical tuning experience, including:

\begin{itemize}
    \item Euclidean distance-based workload similarity: Euclidean distance calculates the straight-line distance between two points in Euclidean space. It is commonly used in machine learning tasks, such as clustering, classification, and regression, to measure the similarity or dissimilarity.  OtterTune~\cite{van2017automatic} applies it to match the target DBMS’s workload  with the most similar workload in its repository based on the performance measurements for the selected group of metrics and then reuses the data from the selected workload to train a GP model on  the target metric for configuration recommendation.
    
    \item Epanechnikov quadratic kernel-based  workload similarity: The Epanechnikov quadratic kernel function, denoted as $K(u)$, is defined as $K(u)=\frac{3}{4}(1-u^{2})$ where the normalized distance $u$ is typically computed as the difference between the coordinates of two points divided by the bandwidth parameter. ResTune~\cite{zhang2021restune} employs it to identify similar historical workloads in its initial tuning stage when there are only limited tuning observation data available for tuning.

    \item Cosine similarity: provides a straightforward and efficient method to measure the similarity between two vectors in a multidimensional space by calculating the cosine of the angle between these vectors. MMOT~\cite{gur2021adaptive} adopts it to determine the model for the target workload by selecting from trained DDPG-based RL models based on workload similarity.
     
    \item Performance surface similarity: works based on the consideration that similar tasks may exhibit comparable trends in terms of where the objective is minimized within the configuration space. It is calculated according to the misranked performance of pairs of given configurations.  ResTune~\cite{zhang2021restune} and OpAdviser~\cite{zhang2023efficient}  identify workload similarity according to the ranking loss of two workloads on a given configuration dataset, utilizing the observed configurations and their corresponding performance as a benchmark. For every pair of configurations, if the ranking of the predicted performance pair does not match the observed pair, the rank loss (count of misranked pairs) is incremented by 1. Finally, the ranking loss for this selected model is defined as the total number of misranked pairings. The smaller the ranking loss, the more similar the selected model is to the target model.
    
\end{itemize}

In the realm of DBMS automatic tuning, knowledge transfer mainly serves two purposes: reducing the need for extensive training data collection and minimizing model training duration. By leveraging this approach, tuning algorithms can implicitly and explicitly utilize extracted knowledge. During the tuning phase, the tuning model can leverage historical data collected from similar workloads (implicitly) or directly apply extracted knowledge, such as ensembling trained models or selecting pre-trained models (explicitly).

\section{Configuration Recommendation}\label{SecRecomendation}

In this section, we embark on an exploration of automatic parameter recommendation methodologies within the realm of DBMSs, categorizing them into four principal domains: Bayesian Optimization, Reinforcement Learning, Neural Network Solutions, and Search-based Solutions. Each classification encompasses a diverse array of methodologies and techniques, each meticulously designed to address specific nuances and challenges encountered within varying DBMS environments and scenarios.

\subsection{Bayesian Optimization}\label{SecBO}
\textbf{Bayesian Optimization} (BO) leverages Bayesian Theorem to direct an efficient and effective search of a global optimization problem. It minimizes/maximizes an objective function $\mathit{f}$ iteratively through adaptive sampling of the search space. BO has two key components: \textit{surrogate model} and \textit{acquisition function}. The surrogate model captures the behavior of the objective function $\mathit{f}$, while the acquisition function guides the selection of the next sample. BO iteratively updates its surrogate model using samples chosen by the acquisition function, ultimately identifying the minimal or maximal value of $\mathit{f}$. 

Surrogate models can be various machine learning models~\cite{bai2023transfer}, such as Random Forest (RF)~\cite{hutter2011sequential} and   Gaussian Process (GP) models~\cite{seeger2004gaussian}, Bayesian neural networks~\cite{snoek2015scalable} or tree parzen estimators~\cite{bergstra2011algorithms}. Among them, GP is a popular surrogate model for objective modeling in BO due to  its expressiveness and well-calibrated uncertainty estimates,  supporting for noisy
observations. GP provides confidence bounds on its predictions to model the objective function with limited samples~\cite{zhang2022towards}. Regarding acquisition functions, popular choices include expected improvement (EI) \cite{jones1998efficient}, probability of improvement (PI) \cite{hoffman2011portfolio}, and GP upper confidence bound (GP-UCB) \cite{rasmussen2003gaussian} or entropy search~\cite{hennig2012entropy}, with EI being the most widely utilized. Sequential Model-based Algorithm Configuration (SMAC) \cite{hutter2011sequential} is a popular variant of BO-based solutions that uses a random forest as its surrogate model, which is known to perform well for high-dimensional and categorical input. It supports all types of variables, including continuous, discrete, and categorical features.

\begin{table*}[htbp]
  \centering
  \caption{\label{BOSolutions}Comparison of BO-based tuning solutions.}
    \resizebox{\linewidth}{!}{
    \begin{tabular}{|p{2.5cm}|l|p{5cm}|p{1cm}|p{3cm}|p{3.7cm}|}
    \hline
    \multicolumn{1}{|l|}{\multirow{2}[0]{*}{\textbf{Work}}} & \multicolumn{1}{c|}{\multirow{2}[0]{*}{\textbf{Acquisition}}} & \multicolumn{3}{c|}{\textbf{Surrogate Model}}&\multicolumn{1}{|c|}{\multirow{2}[0]{*}{\textbf{Highlights}}} \\ \cline{3-5}
          &       & \multicolumn{1}{c|}{\multirow{1}[0]{*}{\textbf{Input features}}} & \multicolumn{1}{c|}{\multirow{1}[0]{*}{\textbf{Model}}}& \multicolumn{1}{c|}{\multirow{1}[0]{*}{\textbf{Initial design}}}& \\ \hline
    LOCAT~\cite{xin2022locat}  & EI-MCMC & Spark SQL configurations (15 selected important paramters) \& data size & GP& 3 samples generated by LHS&-\\ \hline
 
    OtterTune~\cite{van2017automatic}& EI & DBMS configurations (Incremental selection of important parameters ranked by LASSO) & GP&A set of prior observations from the most similar workload&  Increasing the variance of the noise parameter and add a smaller ridge term for each selected configuration\\\hline

    ResTune~\cite{zhang2021restune} & CEI & DBMS configurations ( 14 knobs to optimize CPU, 6 knobs to optimize memory usage, and 20 knobs to optimize the I/O resource) & 3-GP&  A set of prior observations and a small set of known observations on the target task & Ensembling the trained models of historical workloads (including 3 GP models for resource utilization, throughput, and latency, respectively)\\\hline
    REIM~\cite{kunjir2020black} & EI & Spark SQL configurations (6 parameters controlling memory pools across  Container, Application Framework, and JVM) \& 3 empirical memory-related  features & GP&4 samples generated using LHS and 4 noisy observations& Incorporating empirical analytical statistics \\\hline

   ONLINETUNE~\cite{zhang2022towards}& UCB &  DBMS configuration \& Workload dynamicity context (query arrival rate, query composition and three other features related to cardinality estimation and index building /dropping &GP &   Safe configurations set& Augmenting the GP kernel with extra context variables, through an additive kernel  of  context kernel and  configuration kernel\\\hline

   LlamaTune$_{GP}$~\cite{kanellis2022llamatune} & \multicolumn{1}{l|}{\multirow{2}[0]{*}{EI}}  & \multicolumn{1}{c|}{\multirow{2}[0]{*}{Converted features by HeSBO (16)}}  & GP &\multicolumn{1}{l|}{\multirow{2}[0]{*}{10 samples by LHS }}& \multicolumn{1}{l|}{\multirow{2}[0]{*}{Biased sampling}} \\ \cline{1-1} \cline{4-4}
   
   LlamaTune$_{SMAC}$~\cite{kanellis2022llamatune} &  & &  RF (SMAC) & &\\\hline
   
   CGPTuner~\cite{cereda2021cgptuner}& GP-Hedge &  DBMS configurations (15 for MongoDB and 24 for Cassandra) and workload parameters (2-dimension) &GP  &A set of random selected configurations &Designing a multiplication kernel function that works over configuration and workload pairs\\

    \hline
    \end{tabular}%
    }
\end{table*}%

Bayesian optimization-based automatic parameter tuning solutions primarily involve designing the models in terms of the \textit{acquisition function}, \textit{surrogate models}, \textit{model inputs} (i.e., system configurations and configuration constraints), \textit{initial design of samples}, as well as \textit{model update}, to meet the specific tuning objectives. In this section, we provide an overview of state-of-the-art BO-based automatic tuning methods applied in DBMS tuning and BDAFs to describe how BO is adopted in automatic parameter tuning. The summary of different solutions can be seen in Table~\ref{BOSolutions}.

\subsubsection{BO-based solutions}

LOCAT~\cite{xin2022locat} proposes a BO-based approach to ascertain the optimal configurations tailored for Spark SQL applications, with a particular focus on accommodating variations in input data sizes. It begins by employing a workload-level pruning model, known as Query Configuration Sensitivity Analysis (QCSA), to eliminate configuration-insensitive queries within a workload. Meanwhile, it conducts configuration-level pruning using the important configuration parameters (IICP) model through a two-step process: configuration parameter selection (CPS) and configuration parameter extraction (CPE). Subsequently, LOCAT utilizes the BO model along with its proposed Datasize-Aware Gaussian Process (DAGP), which characterizes the application's performance as a distribution of functions of configuration parameters and input data size to recommend the optimal configuration. In its BO-based solution, it employs the Expected Improvement (EI) with a Markov Chain Monte Carlo (EI-MCMC) hyperparameter marginalization algorithm~\cite{snoek2012practical} as the acquisition function, leveraging the proposed DAGP as the surrogate model. Regrading its initial design, LOCAT incrementally constructs the GP model, initially starting with three samples generated through Latin Hypercube Sampling (LHS). Each sample includes information on execution time, configuration, and the corresponding data size. Following each execution, the updated GP model assists BO in selecting the next candidate configuration, aimed at minimizing the execution time of a Spark SQL application.

OtterTune\cite{van2017automatic} proposes to initiate the BO-based solution for DBMS configuration tuning through the observation data from similar workloads. Specially, it first maps unseen DBMS workloads to previously observed workloads by measuring the Euclidean distance between workloads and identifies the most similar workload in its repository based on DBMS runtime metrics. In the next step, OtterTune utilizes the selected similar workload observation data to train a GP surrogate model and updates it with new observations of the target workload. Meanwhile, in order to improve the model's generalization ability on the target workload, OtterTune incorporates some noisy data processing techniques, including increasing the variance of the noise parameter for all points in the GP model that OtterTune has not yet tried for this tuning session and adding a smaller ridge term for each configuration that OtterTune selects. Regarding the acquisition function,  OtterTune adopts EI and it consistently selects configurations with the greatest EI. Meanwhile \textit{Gradient descent}~\cite{hastie2009elements} is applied to identify the local optimum on the surface predicted by the GP model, using data comprising the top-performing configurations completed in the current tuning session and randomly selected configurations.


ResTune~\cite{zhang2021restune} introduces a Constrained Bayesian Optimization (CBO) solution, which extends BO into an inequality-constrained setting to optimize resource utilization without violating service-level agreement constraints during the DBMS configuration tuning process. Specifically, it models each DBMS tuning task with a multi-output GP model which is composed of three independent GP models, representing resource utilization, throughput, and latency requirements. Meanwhile, ResTune introduces a Constrained Expected Improvement (CEI) as its acquisition function. ResTune redefines the best configuration point as the one with the lowest resource utilization while satisfying throughput and latency constraints. CEI calculates the expected improvement of a specific configuration over the best feasible point so far, weighted by the probability that both throughput and latency satisfy the constraints. Moreover, to expedite training, ResTune adopts an ensemble framework to learn from experience by assembling pre-trained learners (GP models) from previous tuning workloads in a weighted manner. In the initial tuning stage, it assigns static weights to previous models based on the workload embedding features. With sufficient observations collected, ResTune assigns dynamic weights according to the ranking loss based on their performance surface similarity.

ReIM~\cite{kunjir2020black} introduces a BO-based solution tailored for the optimization of memory-based analytics applications. In its framework, it utilizes GP as the surrogate model with EI as the acquisition function. In terms of inputs for the surrogate model, ReIM considers parameters encompassing both the application configuration and a set of empirical analysis statistics related to the objectives outlined by RelM. When searching, a combination of random sampling and standard gradient-based search~\cite{Jonas1989} is carried out to find the highest EI.

ONLINETUNE~\cite{zhang2022towards} provides a BO-based solution for online tuning for DBMSs, addressing safety and dynamicity concerns during the tuning process by adding a dynamic context to the tuning task and narrowing down the search space to a safe configuration search subspace. Specifically, it augments the surrogate model GP with dynamic context variables, including workload dynamicity characteristics based on query arrival rate and query composition, as well as the underlying data distribution, in conjunction with DBMS configurations from historical observations. Additionally, it constructs an additive kernel by incorporating a linear kernel to model the dependencies among contexts and a Martin kernel to capture the nonlinear performance associated with configurations, premising that performance correlations emerge when configurations and contextual factors exhibit similarities. ONLINETUNE adopts the Upper Confidence Bound (UCB) \cite{srinivas2009gaussian} as its acquisition function, aiming to improve the mean estimate of the underlying function and decrease the uncertainty at candidate configurations to the maximum. Moreover, it unifies two sampling criteria: UCB method, which selects configurations with maximal confidence interval upper bound safety, and the Highest Uncertainty Boundary (HUB) method, which selects safe configurations at the boundary of the safety set with the highest uncertainty, according to the epsilon-greedy policy~\cite{mnih2013playing}.

Regrading its safe configuration search subspace, it first constructs a candidate configuration set by identifying a safe configuration subspace near the initial safety configuration (default configuration) and adjusts it iteratively, alternating between a hypercube and a line region towards the global optimum. It then discretizes the adapted subspace to build the candidate set. After that, the safety of each candidate is assessed based on the confidence bounds of the contextual GP (black-box knowledge) and the existing domain knowledge (white-box knowledge, e.g., MysqlTuner) to further remove some unsafe or unpromising configuration candidates. Finally, to address the inherent cubic computation complexity of the GP model, it employs the DBSCAN clustering algorithm~\cite{ester1996density} to classify historical observations into different clusters based on their context and trains a GP model for each cluster. Then, SVM is applied to learn the boundary for each model, mapping each new observation to its respective GP model.

LlamaTune~\cite{kanellis2022llamatune} introduces a framework aimed at leveraging domain knowledge to enhance the sample efficiency of existing BO-based automatic DBMS configuration solutions and RL-based solutions. Initially, it employs a randomized projection technique, HeSBO, to project the configuration space from all documentations into a lower-dimensional subspace. Subsequently, the tuning process is initialized within this constructed subspace. In the context of BO-based solutions, LlamaTune adopts two BO-based frameworks: SMAC~\cite{lindauer2022smac3} and  GP-based BO~\cite{li2021openbox}. Initially, a space-filling sampling method (e.g., LHS) is employed to generate an initial set of points for training the surrogate model. The model then suggests a candidate point maximizing the acquisition function EI. Furthermore, LlamaTune implements various strategies for knob selection and knob discretization.  LlamaTune introduces a novel biased-sampling approach, assigning a higher probability to special values such as (i) disabling certain features, (ii) inferring a knob’s value based on another knob, or (iii) setting a knob’s value using an internal heuristic or predefined value during the random initialization phase. Additionally, it employs bucketization of the knob value space at fixed intervals, limiting the number of unique values any configuration knob can take up to a threshold to reduce the configuration space. 
  
CGPTuner~\cite{cereda2021cgptuner} provides a BO-based online tuning solution for DBMSs, considering the IT stack diversity (the operating systems and Java virtual machine) to which the system is exposed. It addresses the IT stack diversity with the Contextual Gaussian Process Bandit Optimization framework, which is a contextual extension to the Bayesian Optimization framework. Specifically, it utilizes GPs as surrogate models, a multiplication combination of two Mat\'{e}rn 5/2 kernels to model the configuration space and the workload space. For the acquisition function, it follows the GP-Hedge approach~\cite{hoffman2011portfolio}, which, instead of focusing on a specific acquisition function, adopts a portfolio of acquisition functions governed by an online multi-armed bandit strategy, aiming to progressively select the best acquisition function according to its previous performance at each tuning iteration. Its workload characterisation is conducted by an external workload characterisation module, such as  the workload characterisation  provided by Ottertune~\cite{van2017automatic}, Peloton~\cite{pavlo2017self} or even a  a reaction-based characterisation used in compiler autotuning~\cite{cereda2020collaborative}

\subsubsection{Summary} In Bayesian Optimization-based solutions, a critical aspect lies in the careful selection and design of tailored components for BO models, encompassing the acquisition function, surrogate model, and model initialization, among others. Table~\ref{BOSolutions} provides a comparative analysis in these dimensions. Concerning the acquisition function, a prevalent choice among researchers is EI,  and some researchers also tried UCB and GP-Hedge. In the realm of surrogate models, GP emerges as a prominent selection, while SMAC exhibits capabilities in supporting diverse parameter types (continuous, categorical, and conditional) and has demonstrated efficacy in managing the intricacies of high-dimensional and heterogeneous configuration spaces. In terms of sample design, practitioners and researchers can try constructing samples from different aspects, including established techniques like random sampling and LHS, alongside the potential for tailoring samples to specific workload patterns or user-provided sample sets. Moreover, practitioners or researchers can also look into refining kernel designs and employing noise data processing strategies within the ambit of model design.

Owing to the training complexity, another crucial consideration is to accelerate the tuning process from data collection and model training. For each configuration execution, acceleration can be achieved through feature pruning, encompassing workload-level pruning and configuration-level pruning. In terms of model training to enhance the coverage process, BO-based solutions are designed to adopt appropriate sampling techniques and transfer knowledge from experience. Regarding task workload modeling, the task workload can be modeled by training a GP model based on historical tuning observations or by building separate GP models for other workloads and ensembling them together for the target task. Currently, extensive research has been conducted on determining the weight in ensembling GP models in Bayesian optimization~\cite{bai2023transfer,schilling2016scalable,wistuba2016two,wistuba2018scalable,feurer2018scalable,li2022transbo,golovin2017google}.

\subsection{Neural Network}

Neural Network (NN)-based Solutions for automatic tuning leverage the power of artificial neural networks  to model the complex relationships between configuration parameters and system performance, enabling efficient tuning without the need for exhaustive search or manual intervention.

\subsubsection{NN-based solutions}
Tan et al. propose iBTune~\cite{tan2019ibtune}, individualized buffer tuning, to automatically reduce the buffer size for any individual database instance while still maintaining the quality of service for its response time. It leverages information from similar workloads to determine the tolerable miss ratio of each instance. Then, it utilizes the relationship between miss ratios and allocated memory sizes to individually optimize the target buffer pool sizes based on a large deviation analysis for the LRU caching model. To provide a guaranteed level of service level agreement (SLA), it designs a pairwise deep neural network (DNN) that predicts the upper bounds of the request response time by learning features from measurements on pairs of instances. Specifically, in the training phase, the NN network takes input of the performance metrics (logical reads, I/O reads, queries per second (QPS), CPU usage, miss ratio) of the target database instance and a similar database instance, the response time (RT) of the similar instance, and the encoding of the current time. The output of the network is the RT. In the testing phase, the previously observed sample of the target is selected as the similar instance. If the predicted RT meets the requirement, the computed new buffer pool size is sent for execution.

Authors in~\cite{van2021inquiry} proposed Ottertune-DNN by adopting the OtterTune framework~\cite{van2017automatic} and replacing its gaussian process model with a neural network model. Same to Ottertune~\cite{van2017automatic}, the raw data for each previous workload is collected and compared with the target workload. Data from the most similar previous workload are then merged with the target workload data to train a neural network model, which replaces the GP model in the original OtterTune. Finally, the algorithm recommends the next configuration to run by optimizing the model. The neural network model is a stacked 2-layer neural network with 64 neurons connected using ReLU as the activation function, and there is one dropout layer among the two hidden layers. Additionally, Gaussian noise is added to the neural network parameters to control the trade-off between exploration and exploitation, with increased exploitation throughout the tuning session achieved by reducing the scale of the noise.

LITE~\cite{lin2022adaptive} achieves configuration tuning for Spark applications by leveraging the power of NN to analyze the application code semantics as well as scheduling information during its execution. Its framework consists of an offline training phase to estimate the performance of a Spark application for a given knob configuration, and an online recommendation phase to recommend appropriate knobs for a given Spark application. In the offline training phase, LITE proposes a code learning framework, Neural Estimator via Code and Scheduler representation (NECS), which extracts stage-level code and Directed Acyclic Graph (DAG) Scheduler as training features. These features are then concatenated with configuration features (knob values), data features, and computing environment features as inputs to predict execution time by an MLP (Multi-Layer Perceptron) module. In its online recommendation phase, LITE introduces an Adaptive Candidate Generation module to initially construct a configuration set. This process involves utilizing a Random Forest Regression (RFR) model to ascertain the center of the promising knob range, which is determined based on factors such as the input data size and the specific application under consideration. Subsequently, the module establishes the knob searching boundary, derived from the input data. Once these boundaries are defined, the module randomly samples a small number of candidates from the selected search space. Finally, it evaluates the performance of each candidate configuration using the NECS model and recommends the knob values associated with the best-estimated performance. Meanwhile, NECS designs an Adaptive Model Update model to    periodically  fine-tunes the NECS model through adversarial learning   when a predefined batch of new instances are collected.

\subsubsection{Summary}  Neural network solutions offer distinct advantages in capturing complex patterns and relationships between variables, particularly adept at handling large and high-dimensional data spaces. Conversely, Bayesian Optimization can incur computational overhead as the parameter space expands, owing to its reliance on surrogate models such as Gaussian Processes. Regarding training data, Bayesian Optimization is renowned for its data efficiency, rendering it suitable for scenarios characterized by limited historical data availability. In contrast, neural network-based solutions necessitate a more substantial amount of data for training, yet they can yield more precise predictions once trained. However, it is worth noting that they may entail a significant time cost in cloud database tuning due to the acquisition of training data.

\subsection {Reinforcement Learning} 

 Reinforcement Learning (RL) is a subfield of machine learning where an agent learns to make sequential decisions by interacting with an environment. In RL, the agent, often represented as an algorithm or program, seeks to develop a strategy or policy that enables it to maximize a cumulative reward over time. The learning process involves the agent navigating through different states, taking actions based on the policy (a mapping from perceived states of the environment to actions to be taken when in these states), receiving feedback in the form of rewards, and updating the policy. The ultimate goal is to obtain an optimal policy by learning from experiences.  Currently, RL has gained more popularity in various areas, including games~\cite{gelly2007combining, mnih2015human}, robotics~\cite{deisenroth2011learning, peters2008learning}, natural language processing~\cite{li2017end, johnson2017google, bahdanau2016actor}, healthcare~\cite{ling2017diagnostic, yu2021reinforcement}, finance~\cite{brandt2005simulation, moody2001learning}, business management~\cite{theocharous2015ad, rolf2023review}, transportation~\cite{haydari2020deep, qian2019deep}, etc.

According to the methods used to determine the policy, RL can be classified into two three categories: (1) \textit{Value-based RL}, which aims to learn the value or benefit (commonly referred to as Q-value) of all actions and selects the action corresponding to the highest value. Among them, Q-Learning~\cite{watkins1992q} is a classic example of a value-based algorithm where the agent learns a Q-value for each action in each state. The optimal policy is then derived from these learned Q-values. (2) \textit{Policy-based RL} takes a different approach by directly learning the optimal policy. Typically, a parameterized policy is chosen, whose parameters are updated to maximize the expected return using either gradient-based or gradient-free optimization ~\cite{arulkumaran2017deep}. In comparison to value-based algorithms, policy-based approaches are effective for high-dimensional or continuous action spaces and can learn stochastic policies, which are more practical than deterministic policies in value-based RL.

Currently, the \textit{actor-critic} framework has gained popularity as an effective means of combining the benefits of policy-based approaches with value-based approaches. In this framework, the `actor' (policy) estimates the target policy and take an action under a specific state, while the `critic' network (value function) approximates the state-value function to evaluate the current policy. One notable development~\cite{arulkumaran2017deep} in the context of actor-critic algorithms is deterministic policy gradients (DPGs)~\cite{silver2014deterministic}, which extend the standard policy gradient theorems for stochastic policies to deterministic policies. 
Following the success of using deep convolutional neural networks to approximate the optimal action-value function in Deep Q-Networks (DQN) ~\cite{mnih2015human}, the Deep Deterministic Policy Gradient (DDPG)~\cite{lillicrap2015continuous} utilizes neural networks within the DPG framework to operate on high-dimensional visual state spaces.

To address the challenge of optimizing performance in large configuration spaces with limited historical data, researchers have proposed RL-based methods. These approaches leverage RL's capability to explore the configuration space through trial-and-error strategies without being constrained by the number of configuration parameters. The main challenge of using RL in
knob tuning is to design the five modules in RL  as shown in Table~\ref{RLforDBMS}. Generally, the DBMS instance is normally considered as the $\mathit{environment}$, and the DBMS configuration is treated as the $\mathit{action}$  in the environment, $\mathit{agent}$ can be seen as the tuning model  which receives $\mathit{reward}$, (DBMS performance, such as throughout $T$ and latency $L$) and state (a set of pre-selected database monitors) from DBMSs and updates the $\mathit{policy}$  to guide how to adjust the knobs for getting higher reward (higher performance). and the DBMS configuration vector represents the state resulting from each action. These configurations will be evaluated based on the tuning performance objective, serving as the reward. Table~\ref{RLComparison} provides an overview of state-of-the-art RL-based configuration tuning solutions applied in DBMS and BDAF.

\begin{table*}[htbp]
  \centering
  \caption{\label{RLforDBMS}RL for DBMS configuration parameter tuning}
    \begin{tabular}{|l|p{12cm}|}
    \hline
    \multicolumn{1}{|c}{\textbf{Terms in RL}}& \multicolumn{1}{|c|}{\textbf{Terms in DBMS}}\\ \hline
    Agent& The RL model, which receives reward (i.e., the performance change) and state from the DB instance and updates the policy to guide how to adjust the knobs for getting a higher reward.\\\hline
    State& State means the current state of the agent, i.e., the DBMS running time internal metrics.\\\hline
    Environment& A DBMS instance.\\ \hline
    Reward&  Reward $r$  is a scalar indicating the objective performance of executing a workload, such as throughput $T$ and latency $L$ or their changes.\\ \hline
    Action& Action comes from the space of knob configurations and represents a knob tuning operation.\\ \hline
    Policy& Policy defines the behavior of the RL model at a certain specific time and environment, which is a mapping from state to action.\\ \hline
    \end{tabular}%

\end{table*}%


\begin{table*}[htbp]
  \centering

  \caption{\label{RLComparison}Comparison of RL-based solutions}

\resizebox{\linewidth}{!}{
    \begin{tabular}{|l|p{3cm}|l|p{4cm}|p{5cm}|}
    \hline

    \multicolumn{1}{|c}{\textbf{Work}}& \multicolumn{1}{|c}{\textbf{State}}&\multicolumn{1}{|c|}{\textbf{Policy}}&\multicolumn{1}{|c|}{\textbf{Reward Function}}&\multicolumn{1}{|c|}{\textbf{Highlights}}\\ \hline
    CDBTune \cite{zhang2019end}&DBMS
internal metrics &DDPG&$r=\alpha  f(\Delta T_{t \to 0},\Delta T_{t\to t-1})+\beta f(\Delta L_{t\to 0},\Delta L_{t\to t-1})
 $&Random Sampling on sample selection\\ \hline
LlamaTune$_{RL}$~\cite{kanellis2022llamatune}&DBMS
internal metrics &DDPG &Same as CDBTune&Configuration space projection, biased sampling\\ \hline
    Qtune~\cite{li2019qtune}& DBMS internal metrics &DS-DDPG&Same as CDBTune&Incorporating query features\\\hline
    MMOT~\cite{gur2021adaptive}&DBMS internal metrics &Multiple DDPG&  $r=\alpha \Delta T_{t \to 0}+\beta \Delta L_{t\to 0} $   &Prioritized  model selection, Gaussian exploration noise\\\hline

    WATuning~\cite{ge2021watuning}& DBMS internal metrics &DDPG with attention & $r=\alpha  f(\Delta T_{t \to 0},\Delta T_{t\to t-1})+\beta f(\Delta L_{t\to 0},\Delta L_{t\to t-1})
 $  or 0  &  State feature importance\\ \hline
    
    HUNTER~\cite{cai2022hunter} &DBMS internal metrics& DDPG with FES &$r=\alpha \Delta T_{t \to 0}+\beta \Delta L_{t\to 0} $&GA sampling, metric compression and knob reduction, and FES on action selection\\ \hline
    DB-BERT~\cite{trummer2022db}& hint-related parameter operations& Double DQN &$ r=|T_{0}-T|$ or -1000&BERT-based manual analysis\\
    \hline
    \end{tabular}%
}
\end{table*}%

\subsubsection{RL-based Solutions}

CDBTune \cite{zhang2019end}  handles the continuous parameter space and unseen dependencies among parameters by adopting DDPG. In CDBtune, the DBMS instance to be tuned is treated as the \textit{environment}, with the DBMS internal metric vector (such as INNODB\_METRICS~\cite{InnoBDMetrics}) representing the \textit{state} from the environment and each DBMS configuration vector as \textit{an action}. Then, the tuning agent generates a configuration and receives a \textit{reward} (consisting of objective performance metrics) after deploying it on the DBMS instance. CDB designs its reward function by considering both throughput and latency across different phases, including the initial tuning state, the previous action, as well as the current action. In this framework, DDPG has a parameterized \textit{policy} neural network that maps the state to the action, and the critic neural network aims to represent the value (score) associated with a specific action and state, guiding the learning of the actor. Meanwhile, CDBtune can learn from past experiences through experience replay in DDPG, which stores past experiences (state, action, reward, next state) in a replay buffer and samples them randomly during the learning process.

LlamaTune$_{RL}$~\cite{kanellis2022llamatune} introduces a framework aimed at leveraging domain knowledge to enhance the sample efficiency of existing BO-based automatic DBMS configuration solutions. Similar to its BO-based solutions, LlamaTune$_{RL}$ employs HeSBO to project the configuration space into a low-dimensional subspace and conducts the same biased sampling knob selection and knob discretization strategies. It integrates the RL framework from CDBTune~\cite{zhang2019end}, utilizing DDPG as its policy function. The converted knob features serve as its states, while throughput/latency metrics are utilized as the reward in the RL framework.

 Qtune~\cite{li2019qtune} provides a query-level parameter tuning system with a RL-based framework to tune the database configurations. QTune first featurizes the SQL queries by considering rich features of the SQL queries, including query type (insert/delete/update/select), tables, and query cost (generated from the query optimizer in terms of scan, hash join, aggregate, etc.). Then QTune feeds the query features into the DRL model to dynamically choose suitable configurations. A Double-State Deep Deterministic Policy Gradient (DS-DDPG) model is applied to enable query-aware database configuration tuning, which utilizes the actor-critic networks to tune the database configurations based on both the query vector and database states.

MMOT~\cite{gur2021adaptive} introduces a multi-model online tuning algorithm employing RL to effectively adapt to dynamic workloads, encompassing fluctuations in system resources, workload diversity, and database size. Initially, it curates a selection of trained DDPG-based RL models derived from historically similar workloads. In each training iteration, the algorithm selects the model with the highest probability of enhancing the reward from the repository and updates it with the current observations. Moreover, it maintains a fresh RL model trained exclusively on the target workload observations. To ensure optimal model selection, models from the repository are pitted against the fresh model. In the event of workload shifts, the most frequently selected model within the workload cycle is retained in the repository. Throughout the iterative process, all RL models employ DDPG as their policy function, utilizing 22 PostgreSQL metrics from three different perspectives to define the action space. Its reward function is degined to incorporate initial throughput and latency, along with current observed performance in each iteration. Meanwhile and it adds Gaussian
action space exploration noise using the Ornstein-Uhlenbeck procesess. Workload characterization is facilitated through the application of a 3-layer autoencoder, enabling both similarity model selection and model shift detection.

WATuning~\cite{ge2021watuning} is a workload-aware auto-tuning system specifically developed to tackle the adaptability challenges posed by changes in workload characteristics within CDBtune. It operates by using a multi-instance mechanism that dynamically employs different trained models to suggest parameter adjustments based on the workload's write/read ratio during runtime. Each instance model is equipped with an attention-based DDPG network, which uses a deep neural network to create a weight matrix capturing the significance of each state characteristic to the current workload. This attention mechanism is  integrated into the actor network, where it collaborates with the critic network to determine the most optimal parameter settings. The reward function is  calculated based on performance improvements observed in throughput and latency in the initial and last time, while also considering deviations from optimal performance and server stability. Meanwhile, it also offers the flexibility to adjust weights for throughput and latency rewards according to user preferences.

HUNTER~\cite{cai2022hunter} proposed an online DBMS tuning system with a hybrid architecture, utilizing samples generated by Genetic Algorithm (GA) to select warm-start samples for the finer-grained exploration of deep reinforcement learning. Subsequently, it employs Principal Component Analysis and Random Forest for metric compression and knob selection, respectively. Deep Deterministic Policy Gradients (DDPG) are then applied to learn and recommend the optimal configuration based on dimension-reduced metrics and selected knobs. Additionally, HUNTER introduced a Fast Exploration Strategy (FES) for DDPG, involving a probabilistic action selection either on the action generated in the current step or on the action that yields the best performance. This strategy forces the model to select configurations with the best performance in the preliminary stage of tuning and encourages exploration based on relatively better configurations to reduce the update time of the learning model.

DB-BERT~\cite{trummer2022db} introduces a RL-based solution for DBMS parameters tuning, leveraging insights extracted from textual sources such as manuals and relevant documents. In its initial data processing phase, DB-BERT employs a finetuned BERT Transformer model to identify and extract tuning hints from textual data, and  prioritized these hints with a heuristic approach, favoring parameters frequently mentioned while ensuring a balanced consideration of hints for each parameter. During the subsequent learning phase, DB-BERT iteratively learns to translate, adapt, prioritize, and aggregate hints to optimize DBMS performance through RF. In each iteration, DB-BERT considers a batch of tuning hints. For each hint, it translates the hint text into a simple equation, assigning a value to a parameter, and then decides whether to deviate from the recommended value (adopt) and assigns a weight to the hint (prioritize). Then, it aggregates them into a small set of configurations, mediating between inconsistent recommendations using hint weights. From the perspective of RL, the environment is defined by workload and system properties. Actions within this environment empower the agent to adapt hints, prioritize between different hints, and aggregate hints for evaluation in trial runs. The performance observed during these trial runs, measured against user-defined benchmarks, serves as a reward signal. This reward signal is then utilized by the Double Deep Q-Networks (DDQN) reinforcement learning algorithm to guide the selection of configurations in future iterations.

\subsubsection{Summary} 
Compared to BO-based solutions and NN-based solutions, RL-based solutions alleviate the difficulty of collecting data for models owing to their reward-feedback mechanism. They learn the tuning strategy through interactions between the database and tuning model, iteratively improving decision-making over time. This makes RL particularly suitable for optimizing performance in large configuration spaces with limited historical data, offering robust performance in complex and uncertain environments.

Existing RL-based DBMS parameter tuning solutions primarily elucidate the model design through four key aspects, as detailed in Table~\ref{RLComparison}: the policy function for identifying optimal knob settings, the reward function for evaluating selected configurations, candidate configuration searching, and model training. To summarize, DDPG stands out as the predominant policy function, given its proficiency in managing continuous parameter spaces. Regarding reward function design, some researchers adhere to the CDBTune reward function, which assesses the performance change of both throughput and latency across various phases. Conversely, others may focus solely on performance alterations during the initial phase or concentrate on a singular performance metric. In terms of searching for candidate configurations, Hunter introduced probabilistic action selection, either on the action generated in the current step or on the action that yields the best performance. Furthermore, researchers propose various methodologies concerning model training, such as sample selection, configuration space transformation, and the integration of supplementary information like tuning experience and state importance into RL models.

\subsection{Search-based Solutions} 

Search-based methods iteratively search for optimal configuration parameters by evaluating the system's performance under different configurations. The core of search-based methods is to design the search strategy to handle the tradeoff between exploration(searching broadly across the configuration space to discover new and potentially promising regions) and exploitation (intensively searching in regions believed to be promising based on the available information).

BestConfig~\cite{zhu2017bestconfig} recommends configuration parameters  through an iterative process consisting of the divide \& diverge sampling for sample collection and the recursive bound \& search for configuration parameter recommendation. Specifically, it first discretizes the parameter ranges for each parameter into several intervals (divide), and then selects a set of samples by considering each interval of a parameter once (diverge). In the recursive bound \& search phase, it starts by searching the point with the best performance among the selected samples. Then, it searches from its surrounding points for better performance (bound). If such point are found, it conducts sampling around the new sample. This optimization process is iteratively conducted until no point with better performance is found.

AutoTune~\cite{bao2018learning} is an automatic parameter tuning system that aims to optimize application execution time on big data analytics frameworks. It executes experiments on both the production cluster and on a smaller-scale testbed to perform more test runs. It consists of three main steps. First, AutoTune executes the application using different data sizes and numbers of machines to build a parametric model. The model is used to select the data and cluster size for running the experiments within a time budget. The second step involves (i) exploration via using latin hypercube sampling (LHS) for generating different configurations to execute and (ii) exploitation via building and employing a Random Forest model for finding promising configurations to execute. In the final step, the best q configurations are used for executing the application on the production and determining the best one. 

UDAO~\cite{song2021spark} considers the tuning problem as a principled multi-objective optimization (MOO) approach that takes multiple, possibly conflicting, objectives, computes a Pareto optimal set of configurations (i.e., not dominated by any other configuration in all objectives), and returns one from the set that best suits the objectives. To be specific, it incrementally transforming a MOO problem to a set of constrained optimization (CO) problems, where each CO problem can be solved individually to return a Pareto optimal point.

 \section{Experimental Setting  \label{SecBenchmarking}}

In the realm of database performance optimization, a diverse array of open-source benchmarks has emerged. These benchmarks encapsulate both data sets and workloads, effectively emulating real-world database systems. Typically, they are categorized into online transaction processing (OLTP) and online analytical processing (OLAP). OLAP prioritizes complex  data analysis and reporting, whereas OLTP specializes in transactional processing and real-time updates.

Typical OLTP benchmarks are listed as follows:

\begin{itemize}

    \item \textit{Sysbench}\footnote{https://github.com/akopytov/sysbench}:  is a popular open source scriptable multi-threaded benchmark tool based on LuaJIT. It provides a set of modular, configurable tests that measure various aspects of system performance, including CPU, memory, file I/O, and OLTP-like database performance (supporting MySQL, MariaDB, PostgreSQL, SQLite, and others). It offers various OLTP workloads designed to simulate different transaction scenarios commonly encountered in online transaction processing systems.  These workloads include read-only, read-write, point-select, update-index, insert, and delete operations, each assessing different aspects of database performance such as concurrency, response times, and efficiency of data manipulation.

    \item \textit{TPC-C}\footnote{https://www.tpc.org/TPC}: This is the current industry standard for evaluating the performance of OLTP systems. TPC-C involves a mix of 5 concurrent transactions of different types and complexity either executed on-line or queued for deferred execution. The database is comprised of 9 types of tables with a wide range of record and population sizes.
	\item \textit{Wikipedia}: This OLTP benchmark is derived from the software that runs the popular on-line encyclopedia. Since the website’s underlying software, MediaWiki, is open-source, the real schema, transactions, and queries used in the live website are therefore available.
	\item \textit{SEATS}: The SEATS benchmark~\cite{stonebraker2012seats} models an airline ticketing system where customers search for flights and make on-line reservations. It consists of eight tables and six transaction types. Approximately 60\% of the transactions are read-only (e.g., customers searching for open seats), while the other 40\% involve creating, updating, and deleting reservation records.
	\item \textit{YCSB}: The Yahoo! Cloud Serving Benchmark (YCSB) ~\cite{cooper2010benchmarking} is modeled after data management applications with simple work- loads and high scalability requirements. It is comprised of six OLTP transaction types that access random tuples based on a Zipfian distribution. The database contains a single table with 10 attributes.
	\item \textit{BenchBase}\footnote{https://github.com/cmu-db/benchbase} (formerly OLTPBench~\cite{difallah2013oltp}) is a Multi-DBMS SQL Benchmarking Framework via JDBC. This benchmark suite offers a standardized set of workloads and metrics to assess the performance and scalability of database systems, supporting transactional Benchmarks such as TPC-C, TPC-H, TATP, SmallBank, SEATS, etc. as well as web-oriented benchmarks such as Twitter, Wikipedia, etc., and feature testing benchmarks including ResourceStresser (RS), YCSB, etc.

\end{itemize}

Typical OLAP benchmarks are listed as follows:

\begin{itemize}

    \item \textit{TPC-H}\footnote{https://www.tpc.org/tpch/}: TPC-H is a standard industry database benchmark with 8 tables.  It consists of a suite of business oriented ad-hoc queries and concurrent data modifications. It is similar to TPC-DS that simulates a decision support system workloads that simulates an OLAP environment where there is little prior knowledge of the queries . It contains 8 tables in 3NF schema and 22 queries with varying complexity.
    
        \item \textit{JOB}\footnote{https://github.com/gregrahn/join-order-benchmark}: Join Order Benchmark (JOB)~\cite{leis2015good} is a comprehensive benchmarking framework, with a set of queries over the Internet Movie DataBase (IMDB), designed to assess cardinality estimation and query optimization strategies. The benchmark encompasses 113 multi-join query instances derived from 33 query templates, tailored to the IMDB dataset. Each query is characterized by a varying number of joins, ranging from 3 to 16, with an average of 8 joins per query.  It has 3.6GB data (11GB when counting indexes) and 21 tables.

    \item \textit{TPC-DS}\footnote{https://www.tpc.org/TPC} containing 99 queries, has been widely used in Spark SQL systems for research and development of optimization techniques ~\cite{chiba2018towards,ivanov2015evaluating,ramdane2018partitioning}. It models complex decision support functions to provide highly comparable, controlled, and repeatable tasks in evaluating the performance of Spark SQL systems.   

    \item The \textit{HiBench}~\cite{huang2010hibench} is a big data benchmark suite that helps evaluate different big data frameworks in terms of speed, throughput and system resource utilizations. It contains a set of Hadoop, Spark and streaming workloads, including Sort, WordCount, TeraSort, Repartition, SQL, etc.  In its SQL related benchmarks it contains three different kinds of queries: Join, Scan, and Aggregation. Join is a query that typically executes in two phases: Map and Reduce. Scan is a query that consists of only a Map operation initiated by the "select" command, which splits the input value based on the field delimiter and outputs a record. Aggregation is a query that consists of both Map and Reduce operations. The Map operation ("select" command) first splits the input value by the field delimiter and then outputs the field defined by the Reduce operation ("group by" command) as a new key/value pair.
\end{itemize}

\section{Related Work}
 Recently, the authors in \cite{zhang2021facilitating} provided an experimental evaluation of several BO-based solutions and RL-based solutions and demonstrated how hyper-parameter optimization algorithms can be borrowed to further enhance database configuration tuning. After that, the authors in \cite{zhao2023automatic} conducted a survey on the state-of-the-art DBMS tuning methods, including heuristic methods, Bayesian optimization methods, deep learning methods, and reinforcement learning methods. They illustrated the automatic tuning pipeline ranging from data preparation to configuration tuning. 
 
 In this study, in addition to elucidating the automatic tuning framework on DBMSs, we provide an in-depth analysis of each component. Specifically, we outline the primary tuning objectives and summarize three main constraints or factors in automatic configuration tuning in DBMS: overhead, adaptivity, and safety. Regarding workload characterization, we present methods for modeling a workload in terms of queries and DBMS runtime metrics. In the feature pruning section, we offer insights into pruning strategies at both the workload and configuration levels. Concerning knowledge transfer, we consolidate existing techniques and present two directions for adopting this knowledge. In the subsequent configuration recommendation section, we provide an overview of existing methods and highlight the available design considerations for each category. Additionally, we include several relevant automatic tuning methods that focus on big data analytics frameworks, which face similar challenges in identifying the optimal configuration within the complex and interdependent configuration space. Finally, in the experiment setting, we summarize popular benchmarks for evaluating DBMS performance.

\section{Discussion and Conclusion}\label{SecConclusion}

This paper presents a comprehensive overview of predominant methodologies utilized in the automatic tuning of parameters within database management systems. The study explores a diverse array of configuration tuning techniques, encompassing Bayesian optimization, Neural network-based approaches, Reinforcement learning methodologies, and Search-based strategies. By systematically dissecting the tuning process into discrete components—comprising tuning objectives, workload characterization, feature pruning, knowledge from experience, configuration recommendation, and experimental settings—this research provides nuanced insights into the strategic intricacies inherent within each phase.

Existing tuning methodologies have undergone extensive investigation into parameter optimization pertaining to DBMS performance, integrating considerations of overhead, adaptivity, and safety concerns. When addressing this tuning task, one essential aspect is workload characterization. The dynamic nature of on-demand cloud applications often necessitates more intricate and varied requirements for the cloud database. These requirements can enrich the application profiling process within the tuning framework, facilitating the optimization of DBMS parameters. Additionally, another essential aspect is data collection and search space reduction. Currently, ML-based solutions, especially for BO and NN solutions, typically require sufficient samples to bootstrap the tuning framework, which can be quite time-intensive. Regarding search space reduction, automatic tuning of DBMS can benefit from innovative research in the hyperparameter optimization problem area, such as distributional variance among source datasets and target datasets~\cite{nomura2021efficient}, as well as search space reduction techniques~\cite{liu2021meta,li2022transfer}. Finally, some other DBMS characteristics can also be considered in the tuning framework, such as database scalability which elucidates performance fluctuations in response to changes in resource capacity, and database elasticity denoting the speed and precision with which a system adapts its allocated resources to varying load intensities. They also emerge as critical considerations in contemporary cloud computing environments~\cite{agrawal2011database,loesing2015design,herbst2013elasticity,papaioannou2017incremental,seybold2019kaa}.

\bibliographystyle{IEEEtran}
\bibliography{Cotuning-sample-base}

\begin{IEEEbiography}[{\includegraphics[width=1in,height=1.25in,clip,keepaspectratio]{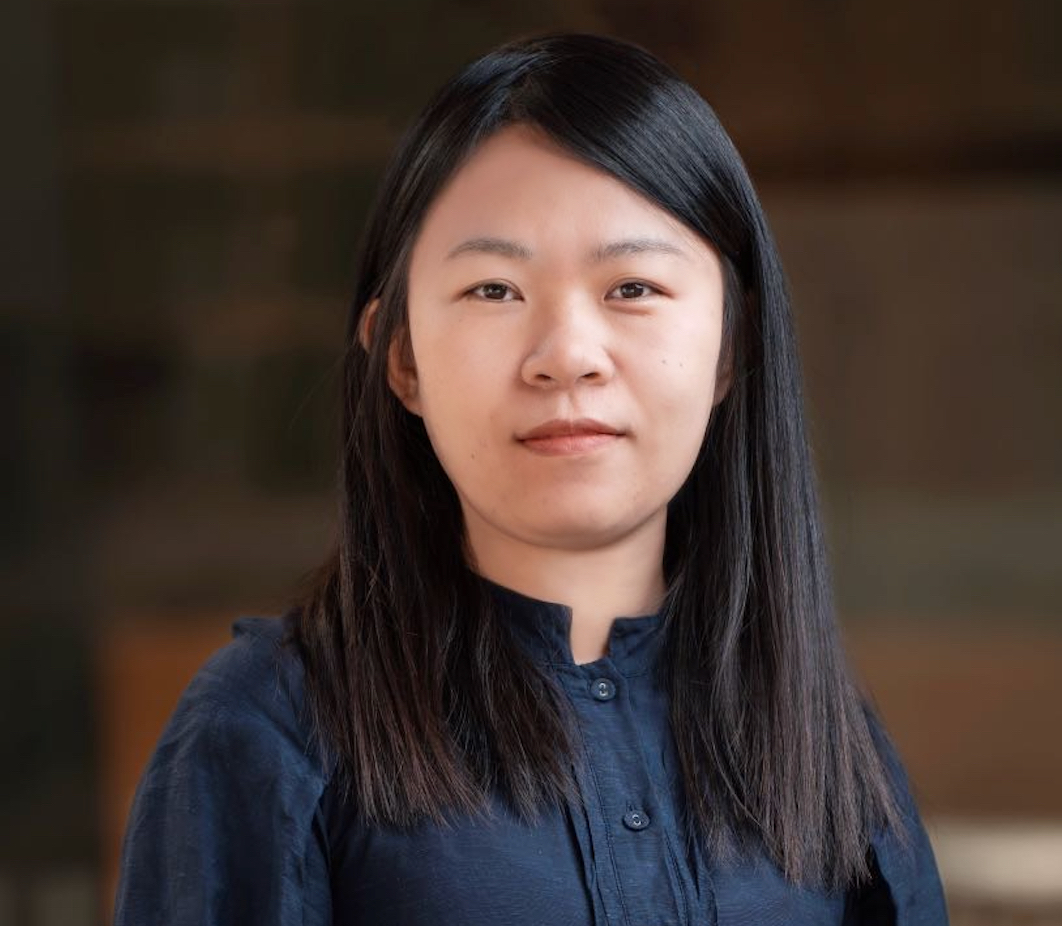}}]{Limeng Zhang} received the B.S. and M.S. and
PhD degrees in computer science from Sichuan University, China, University of Chinese Academy of Sciences, China, and Swinburne University of Technology, Australia, respectively. She
is currently a research fellow with the Centre for Research on Engineering Software Technologies (CREST)
at the University of Adelaide, Australia. Her research interests cover the broad area of Cloud
Computing, Blockchain and Data Mining with special
emphasis on algorithm designs and performance optimization.
\end{IEEEbiography}

\begin{IEEEbiography}[{\includegraphics[width=1in,height=1.25in,clip,keepaspectratio]{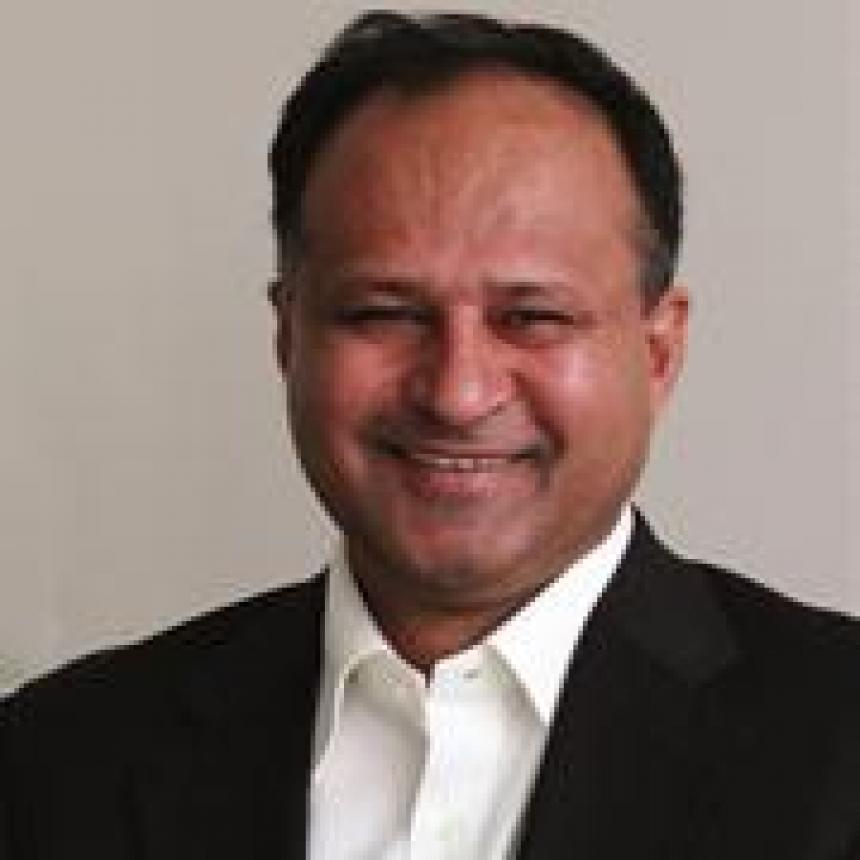}}]{M. Ali Babar}
is a Professor in the School of
Computer Science, University of Adelaide. He is
a honorary visiting professor at the Software Institute, Nanjing University, China. Prof Babar has
established an interdisciplinary research centre,
CREST - Centre for Research on Engineering
Software Technologies, where he leads the research and research training of more than 30
(10 PhD students) members. He leads a theme,
Platforms and Architectures for Cybersecurity as
Service, of the Cyber Security Cooperative Research Centre (CSCRC). Prof Babar has authored/co-authored more than 220 peer-reviewed publications through premier Software Technology journals and conferences. In the area of Software Engineering
education, Prof Babar led the University’s effort to redevelop a Bachelor of Engineering (Software) degree that has been accredited by the
Australian Computer Society and the Engineers Australia (ACS/EA). He coordinates both undergraduate and postgraduate programs of Software Engineering at the University of Adelaide. Prior to joining the University of Adelaide, he spent almost 7 years in Europe (Ireland, Denmark, and UK) working as a senior researcher and an academic.
Before returning to Australia, he was a Reader in Software Engineering with the Lancaster University.
\end{IEEEbiography}

\vfill

\end{document}